\documentclass[fleqn,usenatbib]{mnras}

\usepackage{newtxtext,newtxmath}

\usepackage[T1]{fontenc}

\DeclareRobustCommand{\VAN}[3]{#2}
\let\VANthebibliography\thebibliography
\def\thebibliography{\DeclareRobustCommand{\VAN}[3]{##3}\VANthebibliography}

\usepackage{graphicx}	
\usepackage{amsmath}	
\usepackage{color, xcolor}
\usepackage{cleveref}

\title[Galactic cosmic ray intensity at GJ~436]{The Earth-like Galactic cosmic ray intensity in the habitable zone of the M dwarf GJ~436}

\author[A. L. Mesquita et al.]{
A. L. Mesquita$^{1}$\thanks{E-mail: mesquita@tcd.ie},
D. Rodgers-Lee$^{1}$ and
A. A. Vidotto$^{1}$
\\
$^{1}$School of Physics, Trinity College Dublin, The University of Dublin, Dublin 2, Ireland}

\date{Accepted XXX. Received YYY; in original form ZZZ}

\pubyear{2021}

\begin{document}
\label{firstpage}
\pagerange{\pageref{firstpage}--\pageref{lastpage}}
\maketitle

\begin{abstract}
Galactic cosmic rays are energetic particles important in the context of life. Many works have investigated the propagation of Galactic cosmic rays through the Sun's heliosphere. However, the cosmic ray fluxes in M dwarf systems are still poorly known. Studying the propagation of Galactic cosmic rays through the astrospheres of M dwarfs is important to understand the effect on their orbiting planets. Here, we focus on the planetary system GJ~436. We perform simulations using a combined 1D cosmic ray transport model and 1D Alfv\'{e}n-wave-driven stellar wind model. We use two stellar wind set-ups: one more magnetically-dominated and the other more thermally-dominated. Although our stellar winds have similar magnetic field and velocity profiles, they have mass-loss rates two orders of magnitude different. Because of this, they give rise to two different astrosphere sizes, one ten times larger than the other. The magnetically-dominated wind modulates the Galactic cosmic rays more at distances $< 0.2\,$au than the thermally-dominated wind due to a higher local wind velocity. Between 0.2 and 1\,au the fluxes for both cases start to converge. However, for distances $> 10\,$au, spatial diffusion dominates, and the flux of GeV cosmic rays is almost unmodulated. We find, irrespective of the wind regime, that the flux of Galactic cosmic rays in the habitable zone of GJ~436 (0.2--0.4\,au) is comparable with intensities observed at Earth. On the other hand, around GJ~436\,b (0.028\,au), both wind regimes predict Galactic cosmic ray fluxes that are approximately $10^4$ times smaller than the values observed at Earth.  
\end{abstract}

\begin{keywords}
diffusion -- methods: numerical -- stars: low-mass -- cosmic rays -- planetary systems: GJ~436.
\end{keywords}



\section{Introduction}
Galactic cosmic rays are energetic particles originating from explosive events such as the acceleration of charged particles in supernova remnants \citep{Enomoto2002, Aharonian2004, Brose2020}. These particles are constantly present in the interstellar medium (ISM), and in particular, Galactic cosmic rays are produced within our own Galaxy. There are extra-galactic sources of cosmic rays, but they are only relevant at much higher particle energies \citep{Blasi2014}. These sources are not being considered in the context of this work. 

As Galactic cosmic rays travel through the heliosphere, they interact with the magnetised solar wind, which is known to cause global and temporal variations in the intensity and energy of the cosmic rays  \citep[see review by][]{Potgieter2013}. This phenomenon is known as \emph{the modulation of Galactic cosmic rays.} The modulation of Galactic cosmic rays has been studied extensively in the context of the Sun and Sun-like stars. In this context, several works have investigated the modulation of Galactic cosmic rays at Earth's orbit for different ages \citep{Scherer2002, Scherer2008, Muller2006, Svensmark2006, Cohen2012, Rodgers2020} to understand the possible effects of the cosmic rays during the Earth's lifetime. 

The size of the astrosphere\footnote{An astrosphere is the equivalent of the heliosphere for other stars.} determines how far the Galactic cosmic rays must travel through a magnetised stellar wind. The astrospheric size is determined by the balance between the ISM and stellar wind ram pressures. Thus, the ISM ram pressure (i.e. the ISM properties, such as the density, velocity and ionization fraction) indirectly influences the propagation of Galactic cosmic rays. Some works modelled the response of the astrosphere under different configurations for the ISM conditions around the heliosphere \citep{Scherer2002, Scherer2008, Muller2006} and other astrospheres \citep{Jasinski2020}. \citet{Muller2006} found that the heliospheric structure and size changed and found a wide range of possible heliopause locations varying from 12\, to 402\,au. They also showed that the Galactic cosmic ray spectrum at the Earth's orbit is significantly affected by the ISM conditions, where larger astrospheres cause more modulation of Galactic cosmic rays. From the extreme sizes of the heliosphere, they found approximately three orders of magnitude difference in the Galactic cosmic ray fluxes at the Earth's orbit.

More recently, some works have also studied the modulation of Galactic cosmic rays for a number of M dwarf stars \citep{Sadovski2018, Herbst2020}.  M dwarfs are low mass, low luminous and cool stars. They are especially interesting because their habitable zone, the region where a planet can sustain liquid water on its surface, is closer to the star \citep{Kasting1993, Selsis2007}. This makes M dwarf systems the perfect candidates for transit observations of potentially habitable planets. As a result, exoplanets around M dwarfs are currently the main targets in searching for life outside our solar system \citep{Scalo2007, Tarter2007}. Close-in exoplanets around low mass stars, such as M dwarfs, are currently easiest to observe due to observation bias in our present-day detection technology.

However, a large fraction of M dwarfs remain magnetically active for a longer period of their lives compared to solar-mass stars \citep{West2004, Scalo2007, West2015, Guinan2016}, with the fraction of active M dwarfs being larger for later spectral types \citep[see, e.g.,][]{West2008}. M dwarfs  can generate strong magnetic fields \citep{Morin2010, Shulyak2019}. Strong stellar activity means that the star could have stronger flares \citep{Vida2017, Tilley2019} and coronal mass ejections \citep{Lammer2007, Khodachenko2007}, more high energetic particles \citep{Griebmeier2005} and it could affect the stellar wind \citep{Vidotto2014}. All of these phenomena can affect  planet habitability \citep{Khodachenko2007, Vida2017, Tilley2019}. The longer exposure time to stellar radiation and stellar energetic particles could also affect the planetary atmosphere \citep{Rimmer2013, Rimmer2014, Tabataba2016, Scheucher2018} and climate \citep{Grenfell2013}.

Active M dwarfs generate strong magnetic fields and have higher levels of magnetic activity. These stars should be efficient at accelerating stellar cosmic rays (energetic particles generated by the star). In addition, they have close-in habitable zones and many observed close-in exoplanets. For this reason, stellar cosmic ray fluxes can be expected to dominate over Galactic cosmic rays up to a given energy  around these stars. Some works have investigate the effects of stellar cosmic rays on exoplanets' magnetospheres and atmospheres \citep{Segura2010, Grenfell2012, Tabataba2016, Scheucher2020}, in the habitable zone of M dwarfs \citep{Fraschetti2019} and at Earth's orbit for different ages \citep{Rodgers2021}. Here, we do not focus on very magnetically active stars. Additionally, we do not consider stellar cosmic rays in this work.

In exoplanet atmospheres, cosmic rays can drive the production of prebiotic molecules \citep{Airapetian2016, Barth2020} which are thought to be important for the origin of life. Cosmic rays may have been relevant for the origin of life on Earth and could potentially be relevant for the origin of life on other planets as well \citep{Airapetian2016, Atri2016}. On the other hand, large fluxes of cosmic rays can be extremely harmful for life as we know it \citep{Shea2000}, as they can damage the DNA in cells \citep{Sridharan2016} and possibly cause cellular mutation \citep{Dartnell2011}. However, the majority of cosmic rays do not interact directly with the planet's surface as the surface is protected by an atmosphere and potentially a magnetosphere as well. Works have shown that the flux of cosmic rays at the planetary surface can be reduced by the existence of a magnetosphere \citep{Grenfell2007, Griebmeier2009, Griebmeier2015} and an atmosphere \citep{Griebmeier2016, Atri2020}. \citet{Griebmeier2015} found that the flux of cosmic rays reaching the planetary atmosphere can be enhanced by more than three orders of magnitude if the planet does not have a protecting magnetic field. Additionally, \citet{Atri2020} found that the radiation dose on the planet surface can be reduced by increasing the depth of its atmospheric column density \citep{Atri2013,Atri2017}. Some works also suggest that cosmic rays could affect the Earth's climate through cloud cover \citep{Svensmark1997, Shaviv2002, Shaviv2003, Kirkby2011, Svensmark2017}.

In our present work, we investigate how the winds of M dwarfs can affect the flux of Galactic cosmic rays that penetrate the astrospheres of these stars. In particular, we use results from \citet{Mesquita2020} who modelled the  wind of GJ~436, a moderately active planet-hosting star. In their work the stellar wind was heated and driven by the presence of Alfv\'{e}n waves originating from the base of the chromosphere. The advantage of the Alfv\'{e}n-wave-driven stellar wind model is that it gives a detailed structure of the wind energetics, such as heating (see \Cref{sub:star} for more details). At the same time, it is particularly difficult to observe the winds of M dwarfs, since they have a rarefied wind. Some methods have been proposed to measure the winds of low-mass stars \citep[see review by][]{Vidotto2021}, such as through astrospheric Lyman-$\alpha$ absorption \citep[][and references therein]{Wood2004, Wood2014}, radio emission \citep{Panagia1975, Lim1996, Fichtinger2017, VidottoD2017}, X-ray emission \citep{Wargelin2001, Wargelin2002}, slingshot prominences \citep{Jardine2019} and exoplanet atmospheric escape \citep{Vidotto2017, Kislyakova2019}. These works help to give some constraints on the mass-loss rate for a number of low-mass stars. However, for the majority of objects, some stellar wind parameters are still not fully known, such as the mass-loss rate and terminal wind velocity. For this reason, in our previous work, we varied a number of input parameters. Here, we selected two wind regimes from \citet{Mesquita2020}, a more magnetically-dominated wind and a thermally-dominated wind, to investigate if the wind regime could affect the flux of Galactic cosmic rays in the habitable zone of GJ~436 and at GJ~436\,b. The most important difference between the two wind regimes is that they have mass-loss rates which differ by two orders of magnitude. 

GJ~436 is a very well studied M2.5 dwarf star due to its close proximity at 10.14\,pc \citep{Turnbull2015}. It has a mass of 0.45\,$M_{\odot}$, a radius of $R_\star=0.437\,R_{\odot}$ \citep{Knutson2011} and a rotation period of 44 days \citep{Bourrier2018}. Chromospheric activity indicates that GJ~436 has modest stellar magnetic activity compatible with an old M dwarf \citep{Butler2004}. GJ~436 hosts at least one known exoplanet, GJ~436\,b, at 0.028\,au (about 14.1$\,R_{\star}$), first discovered by \citet{Butler2004}. GJ~436\,b is a warm-Neptune planet with an orbital period of 2.64 days \citep{Butler2004}. The system itself is very interesting, with the planet being observed to lose a substantial amount of its atmosphere \citep{Kulow2014, Bourrier2015, Ehrenreich2015, Bourrier2016}.

Here, we investigate the intensity of Galactic cosmic rays in the stellar system GJ~436. We use the one-dimensional model of cosmic rays transport from \citet{Rodgers2020} to calculate the spectrum of Galactic cosmic rays at different orbital distances in GJ~436's astrosphere. The model is based on the transport equation of \citet{Parker1965}. This paper consists of the following sections: in \Cref{sec:met}, we describe the transport equation that is used to describe the propagation of the Galactic cosmic rays as they travel inside the astrosphere. We also include information about the stellar wind used as an input parameter in our simulations. Our results on the size of GJ~436's astrosphere for the two different stellar wind regimes, the flux of Galactic cosmic rays in the  habitable zone and around the planet's orbit are given in \Cref{sec:res}. Finally, we discuss the important parameters used in our simulations and compare our results with other works in \Cref{sec:dis}, followed by our conclusions in \Cref{sec:conc}.

\section{Galactic Cosmic ray propagation}
\label{sec:met}
\begin{figure}
	\includegraphics[width=\columnwidth]{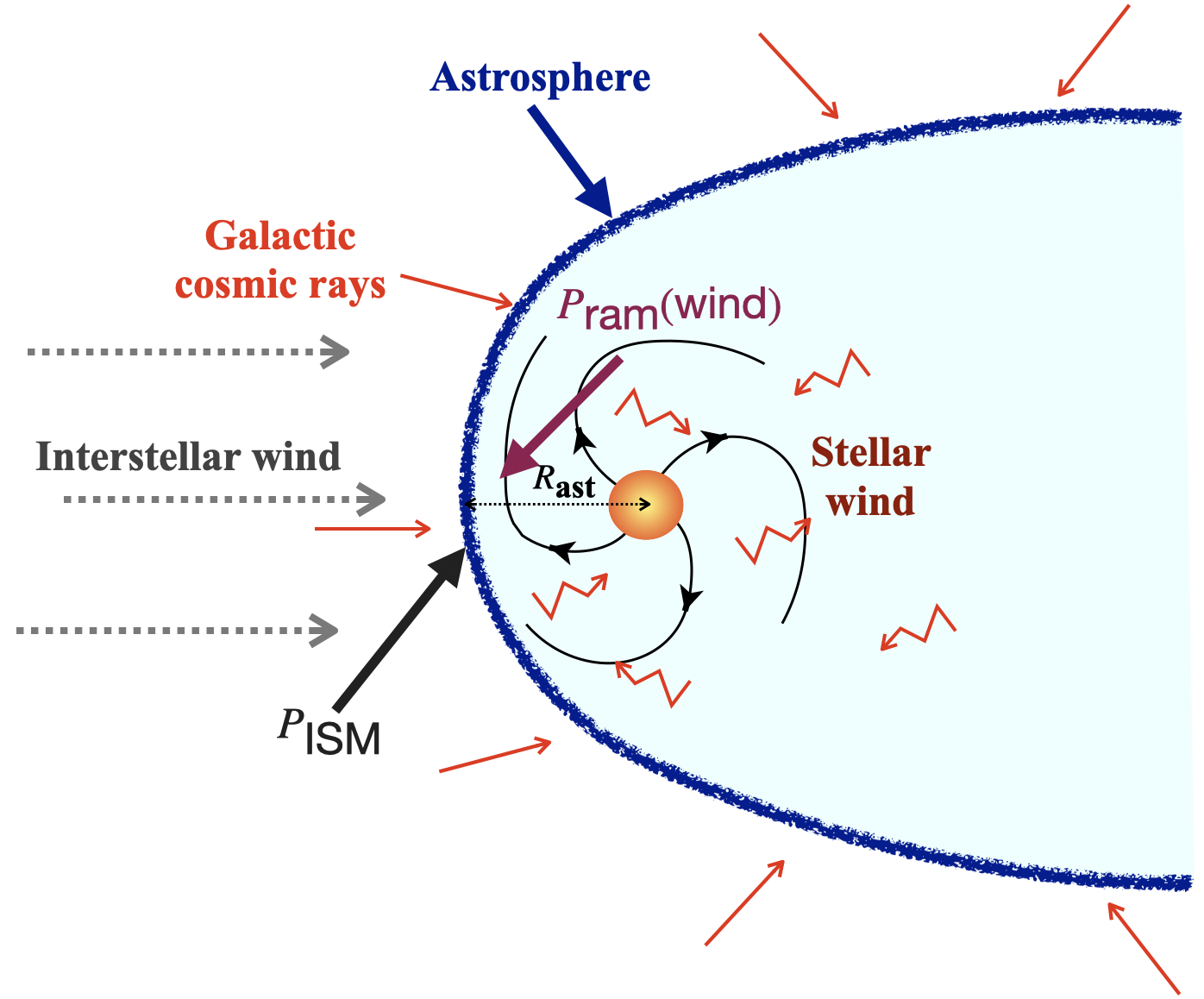}
    \caption{Sketch showing the region dominated by the stellar wind environment (the astrosphere) surrounded by the interaction with the ISM in the inertial frame of the star. The red arrows depict the Galactic cosmic rays propagating into the astrosphere. In red we show the propagation of Galactic cosmic rays as they diffuse through the astrosphere. $P_{\text{ram}}$ is the stellar wind ram pressure and $P_{\text{ISM}}$ is the ISM ram pressure.}
    \label{fig:sketch}
\end{figure}

Galactic cosmic rays travel throughout the ISM. They interact with stellar winds and penetrate the astrospheres of stars, the region of space around stars dominated by the outflow of their magnetised stellar winds. \Cref{fig:sketch} shows a schematic of Galactic cosmic rays propagating inside the astrosphere region. The stellar wind is able to modulate the cosmic rays as they progress inside the astrosphere. Here, we recall that modulation refers to the global variations in the intensity and energy of the cosmic rays as they travel through the stellar wind. In our work, we do not consider the temporal variations. This modulation can be described by the diffusive transport equation of \citet{Parker1965}. The model used in this work is based on the model used in \citet{Rodgers2020} which solves the diffusive transport equation which we describe next.

Here, we numerically solve the one dimensional time-dependent, spherically symmetric transport equation for cosmic rays, given by
\begin{equation}
    \frac{\partial f}{\partial t}=\nabla\cdot(\kappa \nabla f) - u\cdot(\nabla f)+\frac{1}{3}(\nabla\cdot u)\frac{\partial f}{\partial \ln p},
    \label{eq:transport}
\end{equation}
where $f(r, p, t)$ is the cosmic ray phase space density, $u(r)$ is the stellar wind velocity, $\kappa(r, p)$ is the spatial diffusion coefficient, $p$ is the momentum of the cosmic rays and $r$ is the radial distance from the star. The terms on the right-hand side of \Cref{eq:transport} are the spatial diffusion of cosmic rays through the stellar wind, spatial advection and momentum advection, respectively. The spatial advection impedes the propagation of cosmic rays inside the astrosphere and the momentum advection drives the cosmic rays to lower energies, also known as adiabatic losses.

The propagation of cosmic rays through the stellar wind is a diffusive process which depends on the level of turbulence and strength of the magnetic field of the stellar wind. The particles undergo a random walk through the magnetic field lines. From quasi-linear theory \citep{Jokipii1966, Schlickeiser1989}, the diffusion coefficient of the cosmic rays can be expressed as
\begin{equation}
  \frac{\kappa(r,p)}{\tilde{\beta} c}=\eta_0\left(\frac{p}{p_0}\right)^{1-\gamma}r_\text{L},
\end{equation}
where $\tilde{\beta}=v/c$ is the particle speed as a fraction of the speed of light, $p_0=3\,$GeV$/c$, $r_\text{L}=p/eB(r)$ is the Larmor radius of the protons and 
\begin{equation}
    \eta_0=\left(\frac{B}{\delta B_{\text{turbulence}}}\right)^2,
\end{equation}
where $B^2$ is connected with the energy density of the stellar wind large-scale magnetic field and $\delta B_{\text{turbulence}}^2$ to the total energy density in the smaller scale magnetic field turbulent modes. $\eta_0$ represents the level of turbulence in the magnetic field and here we adopted $\eta_0=1$ which sets the maximum value for the turbulence. $\gamma$ is related to the turbulence power spectrum and it defines how the diffusion coefficient changes with energy. Here, we adopt $\gamma=1$ which corresponds to Bohm diffusion and is the same value used by \citet{Svensmark2006}, \citet{Cohen2012} and \citet{Rodgers2020}. With these values, the present day observations of the cosmic ray flux at 1\,au are well reproduced \citep[see Figure A1 from][]{Rodgers2020}. We used logarithmically spaced spatial and momentum grids with 60 grid zones each. The outer spatial boundary for case A is 33\,au and for case B, it is 363\,au (see \Cref{sub:ast}). The inner spatial boundary for both cases is 0.01\,au. The minimum momenta, used in our simulations, is $p_{\text{min}}=0.15\,$GeV$/c$ and maximum momenta $p_{\text{max}}=100\,$GeV$/c$.

In our model, the key parameters necessary for solving the Galactic cosmic ray modulation are: the  Galactic cosmic ray spectrum outside the astrosphere, the stellar wind parameters and the size of the astrosphere (which depends on the ISM parameters combined with the stellar wind parameters), which we will detail in the next subsections. We describe the local interstellar spectrum for Galactic cosmic rays and the fit used in this work in \Cref{sub:lis}. In \Cref{sub:star} we summarise the stellar wind parameters and the wind driving mechanism used. In \Cref{sub:ast} we demonstrate how we calculated the size of GJ~436's astrosphere.

\subsection{Local Interstellar Spectrum (LIS)}
\label{sub:lis}
Outside of the astrosphere Galactic cosmic rays are unmodulated by the stellar wind. For our simulations an unmodulated Galactic cosmic ray spectrum is needed at the outer spatial boundary condition. For this unmodulated boundary condition, we adopt the local interstellar spectrum (LIS).

After crossing the heliopause (the boundary that separates the solar wind and the ISM), Voyager 1 made Galactic cosmic ray observations, which are thought to be unaffected by the solar modulation \citep{Stone2013, Cummings2016}. Using observations of Voyager 1, \citet{Vos2015} developed a model fit to describe the differential intensity of the LIS, $j_{\text{LIS}}$, given by
\begin{equation}
    j_{\text{LIS}}(T)=2.70\frac{T^{1.12}}{\tilde{\beta}^2}\left(\frac{T+0.67}{1.67} \right)^{-3.93}\,\text{m}^{-2}\text{s}^{-1}\text{sr}^{-1}\text{MeV}^{-1},
    \label{eq:jlis}
\end{equation}
where $T$ is the kinetic energy of the cosmic rays in GeV. The differential intensity of cosmic rays can be expressed in terms of the phase space density as $j(T)=p^2f(p)$. In our simulations the LIS is considered to be constant as a function of time. 

\Cref{eq:jlis} describes, by construction, the amount of Galactic cosmic rays at the heliopause at 122\,au and is valid for the solar case. Unfortunately, we do not have observations of the Galactic cosmic ray spectrum outside GJ~436's astrosphere. In-situ measurements of Galactic cosmic rays in locations other than the solar system is not possible. On the other hand, $\gamma$-ray observations of nearby molecular clouds have been used to infer the cosmic ray spectrum for other locations in the Galaxy \citep{Neronov2017, Aharonian2020, Baghmanyan2020}. $\gamma$-ray emission is generated when cosmic rays interact with matter as they travel in the Galaxy. According to \citet{Neronov2017} the inferred cosmic ray spectrum from $\gamma$-ray observations across a region of 1\,kpc in the local Galaxy is in agreement with Voyager measurements at 122\,au. For this reason, it is reasonable to use the LIS measured by Voyager 1 as the outer spatial boundary condition for the simulations of the GJ~436 stellar system.

\subsection{Stellar wind parameters}
\label{sub:star}
The key parameters for our Galactic cosmic ray propagation model are the physical properties of the stellar wind, such as the magnetic field and velocity. To derive such parameters, we use the results of wind simulations from \citet{Mesquita2020}. In their simulations, the wind of GJ436 is heated and accelerated by the dissipation of Alfv\'{e}n waves, analogous to similar processes occurring in the the solar wind (e.g., \citealt{Cranmer2019}; for other Alfv\'{e}n-wave-driven wind models in the context of M dwarfs, see \citealt{Garraffo2016, Mesquita2020, Sakaue2021, Kavanagh2021}). In \citet{Mesquita2020}, the Alfv\'{e}n waves are generated at the base of the wind due to perturbations induced on the magnetic field. The wind is launched at the chromosphere and extends until 0.6\,au (300\,$R_\star$), where the wind has reached its terminal velocity. In their models, the increase in temperature from the chromosphere to the corona occurs naturally. Other models, such as thermally-driven wind models, for instance, cannot model this temperature rise and instead already assume a million Kelvin-temperature wind (e.g., \citealt{Vidotto2014}. See \citealt{Vidotto2021} for a recent review on the winds of low-mass stars).

To investigate the modulation of Galactic cosmic rays around GJ~436 we choose two sets of wind parameters from \citet{Mesquita2020}, which had more than 134 simulations. Note that the different wind models can affect the results of the cosmic ray simulations, we discuss these effects further in \Cref{sec:dis}. Two sets of wind parameters were selected in particular because one is more magnetically-dominated, further referred as `case A', and the other is more thermally-dominated, further referred as `case B'. Although both sets assume the same magnetic field strength at the wind base ($B_0=4$\,G), they have different base densities, which implies that the energy fluxes of the Alfv\'{e}n waves at the wind base are different for each case,  resulting in different radial profiles for the velocity and density, as well as different mass-loss rates.

The first stellar wind, `case A', has a terminal velocity of $u_{\infty}= 1250\,\text{km~s}^{-1}$, a density at the chromosphere of $\rho_0= 4\times 10^{-14}\,\text{g~cm}^{-3}$ and a density at 0.6\,au of $\rho_{0.6\,\text{au}} \sim 6\times 10^{-25}\,\text{g~cm}^{-3}$. The second stellar wind, `case B', has a terminal velocity $u_{\infty}= 1290\,\text{km~s}^{-1}$, a density at the chromosphere of $\rho_0= 3\times 10^{-15}\,\text{g~cm}^{-3}$ and a much higher density at 0.6\,au of $\rho_{0.6\,\text{au}}\sim 7\times 10^{-23}\,\text{g~cm}^{-3}$. The mass-loss rate for Case B is a factor of 125 higher than for case A. \Cref{tab:input} summarises the relevant physical properties of the planet-hosting stellar system, GJ~436. \Cref{tab:cases} shows the stellar wind parameters, the astrosphere sizes and the derived parameters for the modified force field approximation in the habitable zone and at GJ~436\,b orbital distance for each case (see \Cref{sec:HZ}).

In \citet{Mesquita2020}, case A is one of the models in the group called `low-$\beta$ cases' and case B is one of the models in the group called `high-$\beta$ cases', where $\beta$ here refers to the plasma $\beta$ parameter, which is the ratio between the thermal and magnetic pressures. As discussed in \citet{Mesquita2020}, case A is our preferred model for this star, as it predicts an X-ray luminosity that is consistent with that observed for GJ~436 \citep{Ehrenreich2015}.

\begin{table}
 \caption{Properties of the planet-hosting stellar system GJ~436 relevant for this work.}
 \label{tab:input}
 \begin{tabular}{lccc}
  \hline
  Physical parameter & Symbol & Value & Unit \\
  \hline
  Stellar mass$\,^a$ & $M_{\star}$ & 0.452 & $M_{\odot}$\\
  Stellar radius$\,^a$ & $R_\star$ & 0.437 & $R_{\odot}$\\
  Stellar rotation period$\,^b$ & $P_{\text{rot}}$ & 44 & days\\
  Stellar angular speed & $\Omega$ & $1.65\times 10^{-6}$ & $\text{rad~s}^{-1}$\\
  ISM velocity$\,^c$ & $\nu_{\text{ISM}}$ & 81 & km~s$^{-1}$\\
  ISM column density$\,^d$ & $N_{\text{ISM}}$ & $1 \times 10^{18}$  & cm$^{-2}$\\
  ISM average H density$\,^c$ & $n_{\text{ISM}}$ & 0.03 & cm$^{-3}$ \\
  ISM ram pressure$\,^c$ & $P_{\text{ISM}}$ & $3.3 \times 10^{-12}$ & dyn~cm$^{-2}$\\
  Habitable zone & & 0.2 -- 0.4 & au \\
  Semi-major axis$\,^e$ & $a$ & 0.028 & au\\
  \hline
 \end{tabular}
 \newline
    $^a$\citet{Knutson2011}; $^b$\citet{Bourrier2018}; $^c$\citet{Vidotto2017}; $^d$\citet{Bourrier2015}; $^e$\citet{Butler2004}.
\end{table}
 
\begin{table*}
 \caption{Important parameters of each case studied here. We used two Alfv\'{e}n-wave-driven stellar wind models from \citet{Mesquita2020} consisting of a magnetically-dominated wind (case A) and a thermally-dominated wind (case B). The columns are, respectively, the case ID, the terminal velocity, the magnetic field strength and density at the stellar wind base, the magnetic field strength and density at  0.6\,au, the wind mass-loss rate, the Alfv\'{e}n radius, the astrosphere size and the modified force field approximation parameter $\phi$ (see \Cref{sec:HZ}) at the habitable zone (0.2 - 0.4\, au) and at the orbital distance of GJ~436\,b (0.028\,au).}
 \label{tab:cases}
 \begin{tabular}{ccccccccccc}
  \hline
  Case & $u_{\infty}$ & $B_0$ & $\rho_0$ & $B_{0.6\,\text{au}}$ & $\rho_{0.6\,\text{au}}$ & $\dot{M}$ &  R$_A$ & $R_\text{ast}$ & $\phi_{\text{HZ}}$ & $\phi_{\rm GJ~436\,b}$ \\
  & [km~s$^{-1}$] & [G] & [g~cm$^{-3}$] & [G] & [g~cm$^{-3}$] & [$10^{-15}\,M_{\odot}~\text{yr}^{-1}$] & [au] & [au] & [GeV] & [GeV] \\
  \hline
  A & 1250 & 4 & 3 $\times 10^{-15}$ & 3.85 $\times 10^{-5}$ & 6 $\times 10^{-25}$ & 1.2 & 0.08 & 33 & 0.38 - 0.15$\,^*$ & 2.70$\,^\dag$\\
  B & 1290 & 4 & 4 $\times 10^{-14}$ & 3.85 $\times 10^{-5}$ & 7 $\times 10^{-23}$ & 150 & 0.01 & 363 & 0.40 - 0.18$\,^*$ & 2.20$\,^\dag$\\
  \hline
 \end{tabular}
 \newline
    $^*$ The values of $\phi$ in the habitable zone are only well-defined for particle energies above $\gtrsim 200\,$MeV. At lower energies, the analytical expression can overestimate the flux considerably, see \Cref{sec:HZ}.\\
    $^\dag $The values of $\phi$ for GJ~436\,b are only well-defined for particle energies above $\gtrsim 1\,$GeV. At lower energies, the analytical expression can overestimate the flux considerably, see \Cref{sec:HZ}.
\end{table*}

The stellar wind models from \citet{Mesquita2020} end at 0.6\,au (where the wind has reached its terminal velocity) but the astrosphere extends much further out. Thus, we extrapolate the values of $u_r$, $B_r$ and $B_\phi$ beyond 0.6\,au to the astrosphere edge as
\begin{equation}
    u_r(r>0.6\,\text{au})=u_{r,\,0.6\,\text{au}}=u_\infty,
\end{equation}
\begin{equation}
    B_r(r>0.6\,\text{au})=B_{r,\,0.6\,\text{au}}\left(\frac{0.6\,\text{au}}{r}\right)^2,
\end{equation}
\begin{equation}
    B_\phi(r>0.6\,\text{au})=B_{\phi,\,0.6\,\text{au}}\left(\frac{0.6\,\text{au}}{r}\right).
\end{equation}
The velocity is assumed to be constant because the wind already reached its terminal velocity. The radial component of the magnetic field falls with $r^2$ and the azimuthal component falls with $r$, which generates the Parker spiral \citep{Parker1958}.

\begin{figure}
	\includegraphics[width=\columnwidth]{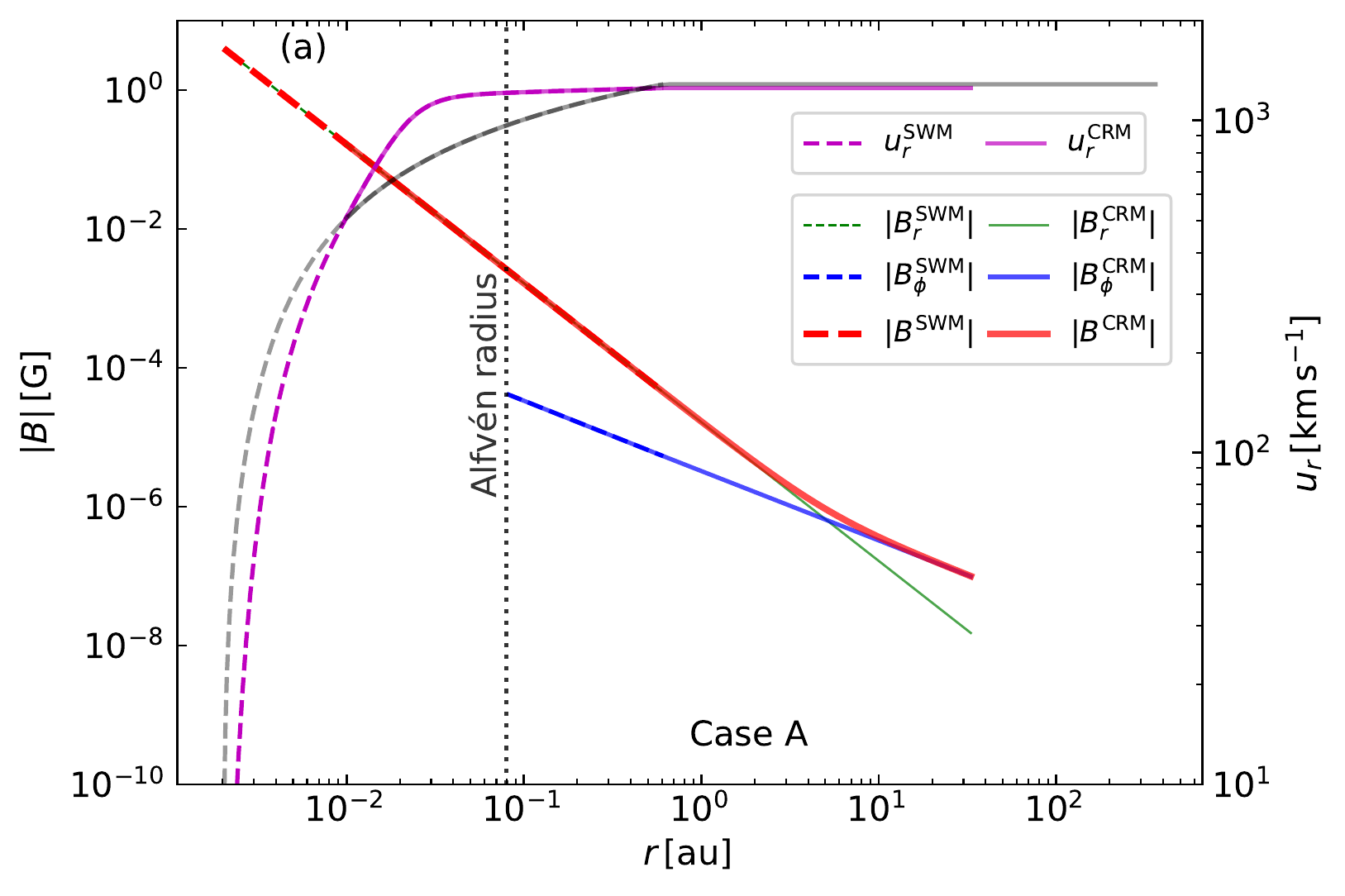}
	\includegraphics[width=\columnwidth]{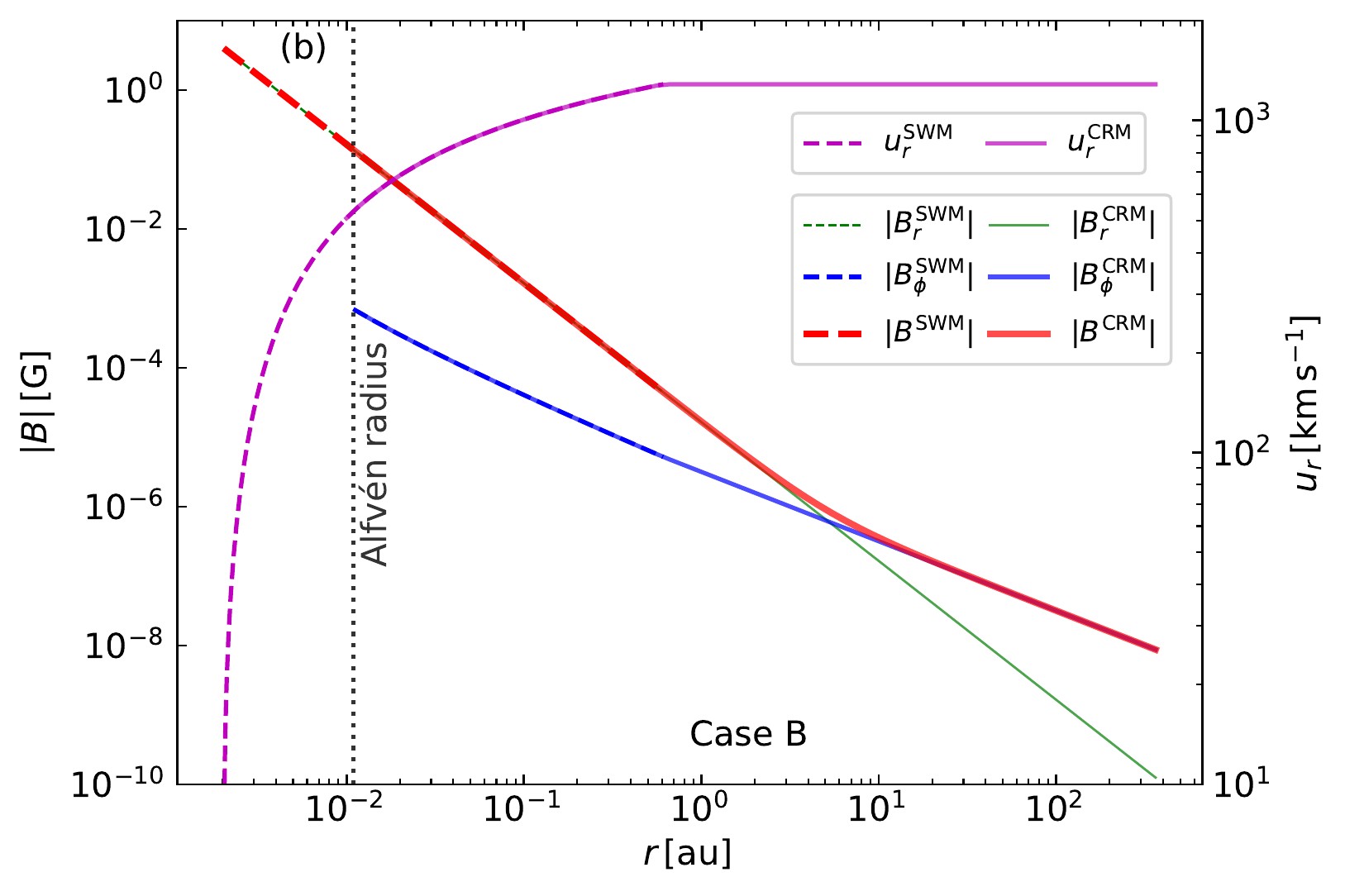}
    \caption{Wind parameters as a function of distance for: a) case A (magnetically-dominated) and b) case B (thermally-dominated). The dashed lines are the output from the Alfv\'{e}n-wave-driven wind simulation and the solid lines are the inputs for the Galactic cosmic ray simulations. The radial velocity profiles are shown in magenta, the radial magnetic field in green, the azimuthal magnetic field in blue and the total magnetic field in red. The grey line on top panel is the velocity profile of case B for comparison. SWM stands for stellar wind model and CRM stands for cosmic ray model.}
    \label{fig:var-cases}
\end{figure}

\Cref{fig:var-cases} shows the wind parameters as a function of distance for the two selected cases. The stellar wind parameter for case A is shown in \Cref{fig:var-cases}-a and for case B in \Cref{fig:var-cases}-b. The dashed lines are the outputs from the Alfv\'{e}n-wave-driven wind simulations and the solid lines are the inputs for the Galactic cosmic ray simulations. The radial velocity profiles are shown in magenta, the radial magnetic field in green, the azimuthal magnetic field in blue and the total magnetic field in red. 

In \citet{Mesquita2020}, the Alfv\'{e}n-wave-driven wind is a 1D MHD simulation which does not take into account rotation, for this reason we do not have a $B_{\phi}$ profile from the stellar wind simulation. Given that at large distances the azimuthal field should dominate, we use the Parker spiral \citep{Parker1958} equation to produce an azimuthal magnetic field component:
\begin{equation}
    \frac{B_\phi}{B_r}=\frac{u_\phi -r\Omega}{u_r},
    \label{eq:spiral}
\end{equation}
where $u_\phi$ is the azimuthal component of the velocity and $\Omega$ is the angular speed of the star. The azimuthal component of the magnetic field only becomes relevant at large distances. Additionally, further out in radius, $u_\phi \ll r\Omega$, and we see from \Cref{eq:spiral} that
\begin{equation}
    |B_\phi|\simeq \frac{r\Omega}{u_r}B_r.
    \label{eq:bphi}
\end{equation}
We use \Cref{eq:bphi} to produce an azimuthal magnetic field for distances beyond the Alfv\'{e}n radius $r>R_A$ (given in \Cref{tab:cases}). The total magnetic field profiles for our simulations consist of the vectorial sum of the radial and azimuthal magnetic field components (see red curves of \Cref{fig:var-cases}). 

\subsection{What is the size of the astrosphere of GJ~436?}
\label{sub:ast}
The outer boundary of the astrosphere is set by the equilibrium between the pressure of the ISM and the stellar wind ram pressure. The size of the astrosphere is relevant in the context of the interaction of Galactic cosmic rays with a stellar wind \citep{Scherer2002, Scherer2008, Muller2006}. This is due to the fact that depending on the size of the astrosphere the Galactic cosmic rays will potentially need to travel a further distance through the magnetised stellar wind and may result in higher levels of modulation, depending on the stellar wind properties.

To calculate the size of an astrosphere, we find the point where the ram pressure of the stellar wind balances the ram pressure of the ISM. The stellar wind ram pressure is given by:
\begin{equation}
    P_{\text{ram}}=\rho u^2,
    \label{eq:ram-wind}
\end{equation}
where $\rho$ is the mass density and $u$ is the velocity of the wind. At large distances, where the wind has reached terminal speed, the density of the stellar wind falls with $r^2$ 
and the stellar wind ram pressure follows the same trend. The distance to the astropause can be estimated as
\begin{equation}
    R_\text{ast}=\left(\frac{P_{\text{ram}}(r=R_{\text{ref}})}{P_{\text{ISM}}}\right)^{1/2}R_{\text{ref}} =\left(\frac{P_{\text{ram}}(r=R_{\text{ref}})}{m_p n_{\text{ISM}}\nu_{\text{ISM}}^2}\right)^{1/2}R_{\text{ref}},
    \label{eq:astro}
\end{equation}
where $R_{\text{ref}}=0.6$\,au is the reference distance adopted in our work where $P_{\text{ram}}$ already falls with $r^2$, $n_{\text{ISM}}$ is the ISM number density of neutral hydrogen, $\nu_{\text{ISM}}$ is the ISM velocity as seen by the star and $m_p$ is the proton mass.

The ISM ram pressure around GJ~436 can be estimated using observations of the ISM column density towards GJ~436 \citep{Bourrier2015}. We follow the same approach as \citet{Vidotto2017}, in the case of GJ~436: we assume the ISM is homogeneous along the line-of-sight of the star and purely neutral. In this way, the ISM density can be estimated as $n_{\text{ISM}}= N_{\text{ISM}}/d=0.03\,\text{cm}^{-3}$ \citep{Vidotto2017}, where $N_{\text{ISM}}=10^{18}\,\text{cm}^{-2}$ is the ISM column density \citep[derived by][]{Bourrier2015} and the distance to GJ~436 is $d=10.14\,\text{pc}$. The density in \Cref{eq:astro} is the total density. If we assume the ISM is purely neutral, then the neutral density and the total density are the same. However, if the ISM is partially ionised \citep[which is the more likely scenario, see, e.g., Table 1 in][]{Jasinski2020}, this means that the total density is larger than the neutral density. As a result, the size of the astrosphere given here (for a purely neutral ISM) is an upper limit. We will discuss the effect of the astrospheric size on the cosmic ray flux further in \Cref{sub:res}. Using Lyman-$\alpha$ reconstruction, \citet{Bourrier2015} found that the ISM absorption towards the line-of-sight of GJ~436, among different possibilities, has a heliocentric radial velocity for the ISM hydrogen most likely associated with the Local Interstellar Cloud (LIC)\footnote{The LIC is the interstellar cloud surrounding the Sun roughly 5--7\,pc across \citep{Redfield2000}.}. Using the ISM Kinematic Calculator \citep{Redfield2008} and the LIC radial velocity from \citet{Bourrier2015}, \citet{Bourrier2015} estimate the heliocentric LIC velocity in the direction of GJ~436. With GJ~436 Hipparcos proper motion \citep{Leeuwen2007} and radial velocity \citep{Bourrier2015}, the heliocentric velocity can be calculated. From these assumptions, \citet{Vidotto2017} derive the ISM velocity as seen by GJ~436 to be $\nu_{\text{ISM}}= 81\,\text{km
~s}^{-1}$ and thus we find an ISM ram pressure of $P_{\text{ISM}}\sim 3.3 \times 10^{-12}\,\text{dyn~cm}^{-2}$.

Using \Cref{eq:ram-wind} and the input parameters (\Cref{tab:input} and \Cref{tab:cases}) we calculate the stellar wind ram pressure to be $P_{\text{ram~A}}(r=0.6\,\text{au}) \sim 9.3\times 10^{-9}\,\text{dyn~cm}^{-2}$ for case A and $P_{\text{ram~B}}(r=0.6\,\text{au}) \sim 1.2\times 10^{-6}\,\text{dyn~cm}^{-2}$ for case B.

Using the values of the ISM ram pressure and the stellar wind ram pressure, we estimate the size of the astrosphere of GJ~436 to be 33\,au for case A and 363\,au for case B. The difference between the size of the astrosphere for the two different cases is due to the wind density for case A being two orders of magnitude smaller than case B. We note that the value calculated in \citet{Vidotto2017} using an isothermal wind model is more similar to the  value we calculate for case A.

\section{Galactic Cosmic Rays in M dwarf systems}
\label{sec:res}
\subsection{Intensity of cosmic rays as a function of the particle's kinetic energy}
\label{sub:res}
\begin{figure}
	\includegraphics[width=\columnwidth]{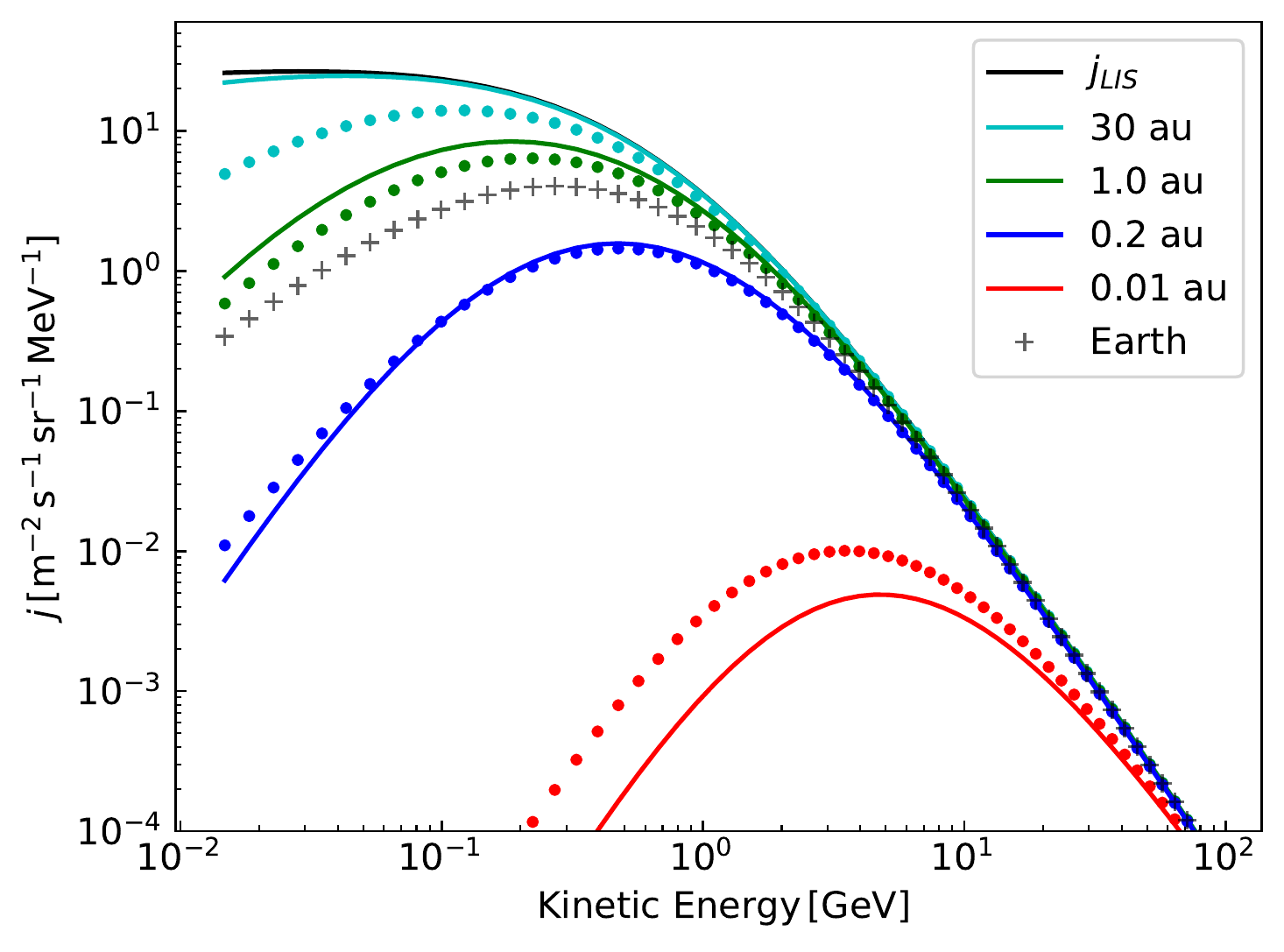}
    \caption{Differential intensity of Galactic cosmic rays at different orbital distances for case A (solid lines) and case B (dotted lines). Smaller orbital distances have a much lower intensity of Galactic cosmic rays compared with larger distances. The black solid line is the LIS and grey crosses are representative of the intensity of Galactic cosmic rays observed at Earth at solar minimum \citep{Rodgers2020}.}
    \label{fig:crs-cases}
\end{figure}

Here we investigate how the modulation of Galactic cosmic rays around GJ~436 is affected by different stellar wind properties. \Cref{fig:crs-cases} shows the intensity of Galactic cosmic rays as a function of the particle's kinetic energy for case A (solid lines) and case B (dotted lines) for different distances. The solid black curve is the LIS given by \Cref{eq:jlis}. Case B has a larger astrosphere which means that the differential intensity of cosmic rays at a given distance will be smaller than for a system with the same wind parameters and a smaller astrosphere. This is noticeable in \Cref{fig:crs-cases} when we analyse the differential intensity of cosmic rays at 30\,au (light blue curves) and 1\,au (green curves) for both cases (at large orbital distances the stellar wind properties for case A and B are nearly identical). In these situations the differential intensity of cosmic rays of case A is  higher than case B. 30\,au is close to the outer boundary for case A, which is why we chose this specific distance. After travelling a short interval, $\sim 3$\,au (solid light blue curve of \Cref{fig:crs-cases}), the cosmic rays have almost the same flux as the LIS for case A. On the other hand, for case B after travelling a larger distance, $\sim 333$\,au (dotted light blue curve of \Cref{fig:crs-cases}), the cosmic rays lose energy but less than an order of magnitude compared with the LIS. This shows that the size of the astrosphere has a small contribution to the modulation of cosmic rays in the case of GJ~436.

The radial velocity profile of case A is slightly higher (by a factor $\lesssim 1.4$) than case B for distances smaller than 0.45\,au (compare the grey and pink lines in \Cref{fig:var-cases}-a). Due to the combination of the velocity profile and the size of the astrosphere, at around 0.2\,au the intensity of cosmic rays is equivalent for both cases (see \Cref{fig:crs-cases} blue lines). For smaller distances ($<0.2\,\text{au}$), case A modulates the cosmic rays more than case B because of the higher velocity of the wind for case A. If the stellar wind velocity is larger it suppresses the flux of Galactic cosmic rays more.

\subsection{Advective and diffusive timescales}
The modulation of Galactic cosmic rays in the system can be understood by analysing the advective and diffusive timescales. The timescales are defined as:
\begin{equation}
    \tau_{\text{adv}}=\frac{r}{u}, \qquad\qquad \tau_{\text{dif}}=\frac{r^2}{\kappa}\propto \frac{r^2}{p/B}\,.
\end{equation}

\begin{figure}
	\includegraphics[width=\columnwidth]{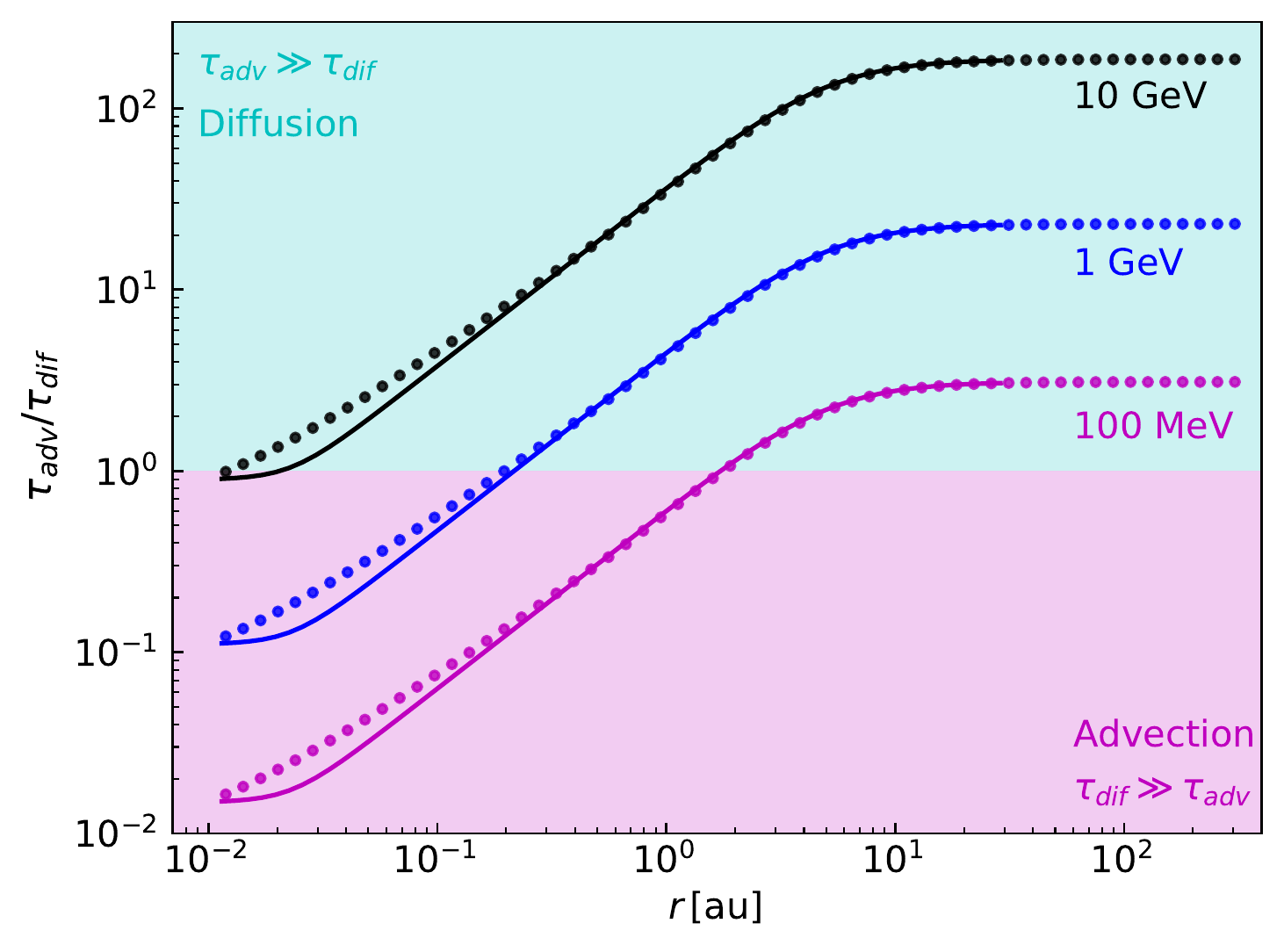}
    \caption{Ratio between advective timescale and diffusive timescale as a function of distance from the star for case A (solid lines) and case B (dotted lines). In the blue region, the dominate physical process is diffusion. In the pink region, Galactic cosmic rays are strongly modulated by advection.}
    \label{fig:timescales}
\end{figure}

The advective timescale depends only on the stellar wind velocity while the diffusion timescale depends on the momentum of the cosmic rays and the magnetic field profile of the stellar wind.
\Cref{fig:timescales} presents the ratio between the advective and the diffusive timescales as a function of distance for case A (solid lines) and case B (dotted lines) for different values of cosmic ray kinetic energy. The pink shaded area ($\tau_{\text{adv}}/\tau_{\text{dif}}<1$) indicates the region where advection dominates and the blue shaded area ($\tau_{\text{adv}}/\tau_{\text{dif}}>1$) indicates the area where diffusion dominates. If diffusion strongly dominates the Galactic cosmic rays experience little (if any) modulation. When timescales are comparable ($\tau_{\text{adv}}/\tau_{\text{dif}}\sim 1$) both effects start to compete and the Galactic cosmic rays start to experience modulation. On the other hand, if advection dominates the Galactic cosmic rays are strongly modulated by the stellar wind. 

The diffusion timescales have the same values for both cases because they have the same magnetic field profile. On the other hand, the advective timescale is slightly different due to the difference in the wind velocity profile of each case. Consequently, the ratio between the timescales is different between both cases which is only noticeable for smaller radii ($r<0.45$\,au).

The timescale ratio can be written as 
\begin{equation*}
    \frac{\tau_{\text{adv}}}{\tau_{\text{diff}}} \propto \frac{p}{ruB}\propto \frac{p}{u}\frac{r^{\alpha}}{r},
\end{equation*}
where $r^\alpha$ comes from the magnetic field profile as $B\propto 1/r^\alpha$. The radial component of the magnetic field falls with $r^2$, hence is best represented by $\alpha \sim 2$. The azimuthal component of the magnetic field falls with $r$ and it is best represented by $\alpha \sim 1$. For distances $r\leqslant10\,$au, $\alpha \sim 2$ (see \Cref{fig:var-cases} green and red curves) and the timescale ratio is $\propto pr/u$. For distances $r\geqslant 10\,$au, $\alpha \sim 1$ (see \Cref{fig:var-cases} blue and red curves) and the timescale ratio is $\propto p/u$. Both trends can be observed in \Cref{fig:timescales}, where the timescale ratio increases with distance for $r\leqslant10\,$au and it becomes constant for $r\geqslant 10\,$au.

For distances smaller than 0.2\,au the timescale ratio of case B is higher than case A, for a given energy. When $r<0.2\,$au, we have $u_A>u_B$ and consequently $\tau_{\text{adv\ B}}>\tau_{\text{adv\ A}}$. Consequently case A modulates the Galactic cosmic rays more than case B in this region. For distances larger than 0.2\,au both cases show a similar value for the timescales.

From \Cref{fig:timescales} we observe that higher energy (10\,GeV, black lines) cosmic rays are unmodulated because their diffusive timescale is much smaller than the advective timescale. In other words, they can diffuse faster in the stellar system than the advective processes can act to suppress them. Low energy cosmic rays (100\,MeV, magenta lines) are more modulated by the stellar wind because their diffusive timescale is much larger than the advective timescale. This means that stellar winds are more effective at modulating low energy Galactic cosmic rays than high energy ones.

The modulation of Galactic cosmic rays is stronger for low energy cosmic rays (< 1\,GeV) and less effective for high energy cosmic rays (> 1\,GeV) because the diffusive timescale depends inversely on the momentum of the cosmic rays. The more energetic cosmic rays are not modulated by the stellar wind which can be observed by the fact that all curves (except red curves) in \Cref{fig:crs-cases} are the same as the LIS (the LIS spectrum is by definition unmodulated) for energies greater than 5\, GeV. In contrast, high energy cosmic rays begin to be modulated at small orbital distances as can be observed by the red curves in \Cref{fig:crs-cases}. In fact, Galactic cosmic rays of all energies that we consider are strongly modulated for small distances which can be explained by the fact that the stellar wind magnetic field is much stronger close to the star leading to large diffusion timescales in combination with a small advective timescale. 

\subsection{The intensity of Galactic cosmic rays in the habitable zone and at GJ~436\,b}
\label{sec:HZ}
\begin{figure}
	\includegraphics[width=\columnwidth]{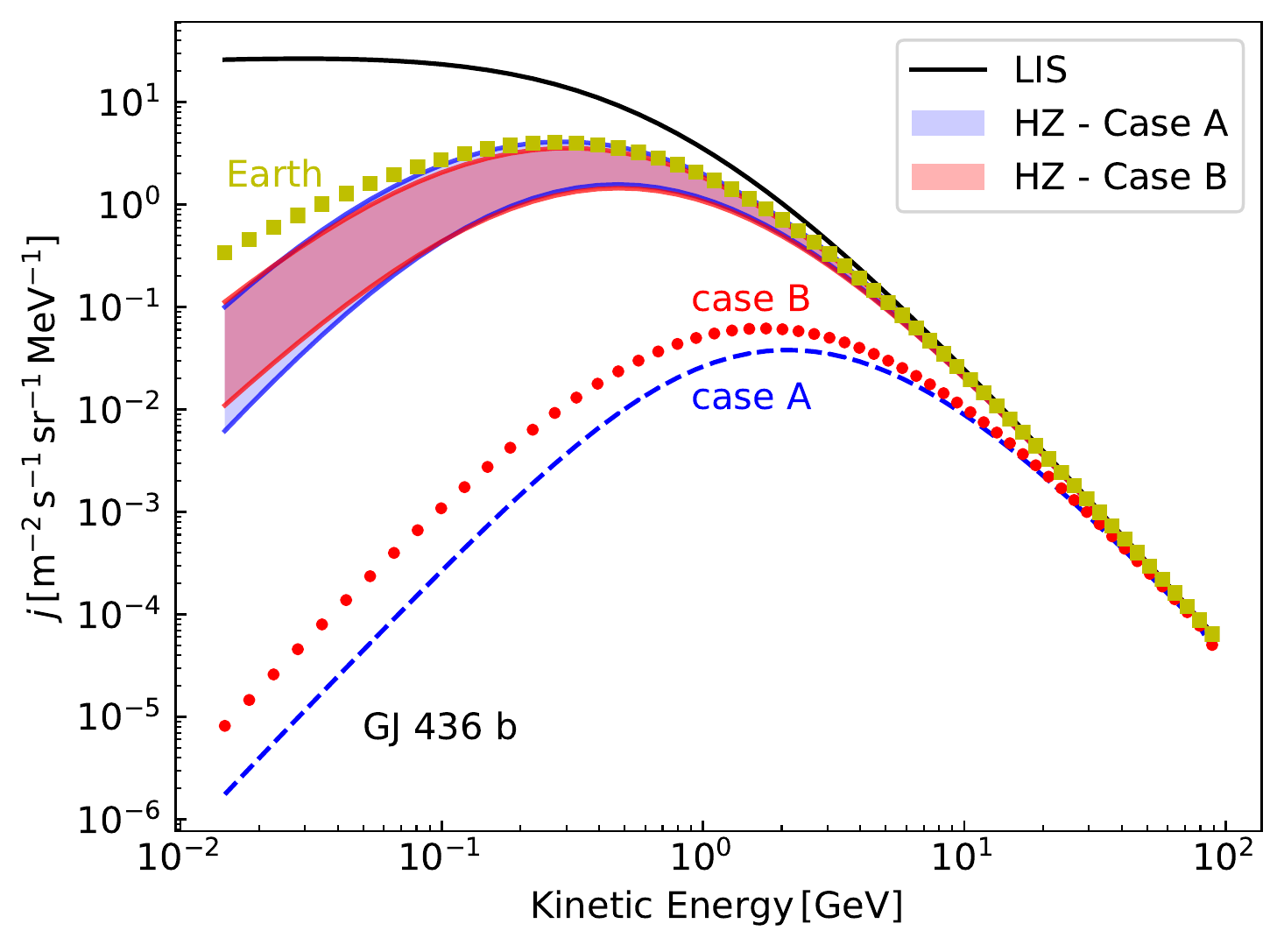}
    \caption{Differential intensity of Galactic cosmic rays in the habitable zone of GJ~436 using case A (blue shaded area) and case B (red shaded area) and at the orbital distance of GJ~436\,b (0.028\,au) for case A (dashed line) and case B (dotted line). The yellow squares is representative of the intensity of Galactic cosmic rays observed at Earth at solar minimum \citep{Rodgers2020}.}
    \label{fig:HZ}
\end{figure}

The region around a star where it is possible for a planet to have liquid water on its surface is called the habitable zone \citep{Kasting1993, Selsis2007}. The differential intensity of Galactic cosmic rays in the habitable zone of GJ~436 (0.2--0.4\,au)\footnote{The habitable zone was calculated based on \citet{Kasting1993} and \citet{Selsis2007} by assuming an Earth-like exoplanet with the same albedo and the same greenhouse effect as the Earth.} is shown in \Cref{fig:HZ} as a function of cosmic ray kinetic energy, where blue is for case A and red for case B. The black line is the LIS and the green points are the flux of Galactic cosmic rays observed at Earth \citep{Rodgers2020}.

Overall, the flux of cosmic rays in the habitable zone of GJ~436 is very similar for both cases, with case B presenting a slightly higher level of modulation than case A. This result is due to a combination of the velocity profile and the astrosphere size of the cases studied here. The similar results we found here is possibly a coincidence, which would not happen if we choose another stellar wind model (see \Cref{sec:dis} for further discussion). In addition, the flux of Galactic cosmic rays in the habitable zone is comparable with values observed at Earth for both cases studied here, even though the habitable zone around GJ~436 is at much smaller distances than the solar habitable zone.

Since GJ~436 hosts a planet, we further investigate the differential intensity of cosmic rays reaching the orbital distance of GJ~436\,b (0.028\,au). The results are shown in \Cref{fig:HZ} as a function of cosmic ray kinetic energy. Because the planet orbits at such a close distance, case A (blue dashed line) shows a higher modulation of cosmic rays when compared with case B (red dotted line). The difference is more pronounced when we analyse low energy particles ($<1\,$GeV). We also present in \Cref{fig:HZ} the flux of Galactic cosmic rays at Earth's orbit (green solid line) as a function of cosmic ray kinetic energy for comparison \citep[see Figure A1 from][]{Rodgers2020}. When compared with the Earth, GJ~436\,b is exposed to far fewer Galactic cosmic rays than the present-day Earth.

We provide here an analytical expression for the Galactic cosmic ray intensity at the habitable zone of GJ~436 and at the orbit of GJ 436\,b. This analytical expression can be used in other works where the flux of cosmic rays is considered an input (e.g., calculating the flux of cosmic rays through planetary magnetospheres and atmospheres). The force field approximation \citep{Gleeson1968} is a theory-based, analytic expression that can be used to describe the modulation of Galactic cosmic rays. This expression depends only on a modulation potential $\phi$. Here, we use a modified force field approximation \citep{Rodgers2020} for the differential intensity of Galactic cosmic rays in the habitable zone of GJ~436, given by:
\begin{equation}
    \frac{j(T)}{E^2-E_p^2}=\beta\left(\frac{j_{\rm LIS}(T+\phi)}{(E+\phi)^2+E_p^2}\right),
    \label{eq:ffa}
\end{equation}
where $E=\sqrt{(pc)^2+E_p^2}$ is the total proton energy, $E_p = 0.938\,$GeV is the proton rest energy and $\phi$ is the average energy lost by a cosmic ray coming from infinity, i.e. the ISM. The modified force field approximation (\Cref{eq:ffa}) has an extra factor of $\beta$ in comparison to the force field approximation which acts to suppress low energy cosmic rays more. The values of $\phi$ which  fit our data well are given in \Cref{tab:cases}. The modified force field approximation can be used to easily reproduce the Galactic cosmic ray spectrum in the habitable zone of GJ~436 and at the orbit of GJ~436\,b. For the former, the modified force field approximation is only appropriate for energies above 200\,MeV, as it overestimated the low energy cosmic ray fluxes by 10 times at 0.015\,GeV, while for the latter, the approximation should only be used above 1\,GeV, as it can overestimate the flux by a factor of up to $\sim 2$ orders of magnitude at 0.015\,GeV.

\section{Discussion}
\label{sec:dis}
When modelling the propagation of Galactic cosmic rays inside a stellar astrosphere there are some key parameters that influence the modulation of Galactic cosmic rays. Here, we discuss how the parameters used in our simulations could possibly affect the Galactic cosmic ray propagation and how our work compares with other work in the literature. We also suggest possible future plans.

From the many stellar wind models available from \citet{Mesquita2020}, there is a set which is compatible with X-ray observations of GJ~436. This set has stellar wind base densities $< 7 \times 10^{-15}\,\text{g~cm}^{-3}$ and stellar magnetic fields varying from 1 to 10\,G, resulting in $\dot{M} < 7.6 \times 10^{-15}~M_\odot$/yr. From their models, we chose only two with similar velocity and magnetic field profiles but that are in a different wind regime and thus have very distinct mass-loss rates (and stellar wind densities). The magnetically-dominated wind (case A) is one of the models that is compatible with observations, as it predicts X-ray luminosities that are below values observed for GJ~436 \citep{Ehrenreich2015}, and gives rise to a mass-loss rate more compatible with other calculated values \citep[see discussion in][]{Mesquita2020}. Case B, on other hand, is not compatible with such observations. In spite of substantial differences in wind mass-loss rates, our two selected wind models gave rise to a similar modulation of Galactic cosmic rays in the habitable zone of GJ~436 which may indicate that our lack of knowledge on the stellar wind regime is not important for this study.

However, there is still the question of whether our choice of models {\it coincidentally} led to similar results. If we had selected another stellar wind model, would it have resulted in a different modulation of Galactic cosmic rays? For instance, a stronger stellar magnetic field leads to longer diffusive timescales, which allows the advective processes to dominate (for a given fixed velocity profile), thus, generating more modulation. A higher stellar wind velocity would also modulate the Galactic cosmic rays further. Additionally, the stellar wind density profile would affect the size of the astrosphere which also affects the modulation of the cosmic rays. \citet{Mesquita2020} gave an upper limit for the mass-loss rate and density for GJ~436, but the magnetic field profile remains unconstrained. One way to constrain the magnetic field of a star is through spectropolarimetric observations \citep[e.g.][]{Morin2010}. However, an observationally-derived magnetic map of GJ~436 is currently not available. A magnetic map would allow us to model the stellar wind in 3D \citep[e.g., ][]{kavanagh2019} and with that the influence of the magnetic field geometry on the cosmic ray propagation could be studied.

The properties of the ISM (such as density, velocity, temperature and ionisation fraction) are also important for the modulation of Galactic cosmic rays, since it is directly connected with the astrosphere size. \citet{Jasinski2020}, following the methodology of \citet{Muller2006}, demonstrated the importance of the ISM conditions by showing how it can affect the Galactic cosmic ray propagation around two stellar systems, Kepler-20 and Kepler-88. Using the ISM density and velocity constrained by \citet{Jasinski2020}, these authors calculated the ISM ram pressure and found that the astrosphere sizes for Kepler-20 and Kepler-88 could be in the range 2--63\,au and 10--270\,au, respectively. \citet{Jasinski2020} found at the orbital distances of Kepler-20f and Kepler-88c that the Galactic cosmic ray fluxes at $\sim 100\,$MeV energies could be $\geqslant 2$ orders of magnitude different due to the astrosphere size variation. Here, we found that the size of the astrosphere did not strongly affect the modulation of Galactic cosmic rays. Our astrosphere size varied due to the different stellar wind conditions rather than the ISM conditions. In our work, an astrosphere ten times larger does not significantly affect the flux of cosmic rays for most radii. At 30\,au, for instance, case B has less than half of the flux of cosmic rays with energy of 0.15\,GeV in comparison with case A at the same distance. The reason for the discrepancy between the work presented in \citet{Jasinski2020} and our work is related to a different formulation of the diffusion coefficient for the cosmic rays. In contrast with our work, \citet{Jasinski2020} use a spatially constant diffusion coefficient. In our work, the diffusion coefficient scales inversely with the magnetic field, similar to \citet{Herbst2020}. We cannot point out which formulation for the diffusion coefficient is more appropriate for GJ 436, or the systems considered in \citet{Jasinski2020}, because the spatial variation of the turbulence in these systems is unknown. Our prescription for the diffusion coefficient is based on observations in the solar system. When applying it to the GJ~436 system, we assume that the spatial variation of the turbulence is similar in both systems.

\citet{Herbst2020} calculated the flux of Galactic cosmic rays throughout the astrosphere of three M dwarf stars (V374 Peg, Proxima Centauri and LHS 1140). They found at the orbital distances of Proxima b and LHS 1140\,b that the Galactic cosmic rays were not significantly modulated. On the other hand, the Galactic cosmic rays are strongly modulated in the V374 Peg system.\footnote{In Fig.~3 of \citet{Herbst2020} the units should read as $\mathrm{m^{-2}\,s^{-1}\,sr^{-1}\,MeV^{-1}}$, instead of $\mathrm{m^{-2}\,s^{-1}\,sr^{-1}\,GeV^{-1}}$ (Herbst, private communication).} Our values for GJ~436\,b lie between these two extremes. LHS 1140 has a very small astrosphere of 11.3\,au, a slow stellar wind and a weak magnetic field which all contribute to the lack of significant Galactic cosmic ray modulation in this stellar system. Proxima Centauri has a weak magnetic field at 1\,au in comparison to the solar value at the same distance which leads to larger diffusion coefficients resulting in less modulation of the Galactic cosmic rays. In comparison, V374 Peg has an astrosphere 8500\,au in size and a strong magnetic field which leads to the strong modulation of Galactic cosmic rays. These results show the variety between the different M dwarfs when studying the propagation of Galactic cosmic rays. One important point to highlight is that \citet{Herbst2020} used a different prescription for the diffusion coefficients with $\gamma=5/3$ (corresponding to Kolmogorov-type turbulence), normalised at 1\,GeV/c. This $\gamma$ value affects the modulation at all energies. Cosmic rays with momentum $<1\,$GeV$/c$ are less modulated, in comparison with our adopted value of $\gamma=1$. Cosmic rays with momenta $>1\,$GeV$/c$ suffers less modulation, in comparison with $\gamma=1$ (due to the normalisation occurring at 1\,GeV$/c$). An additional point to note is that \citet{Herbst2020} used the ISM properties from outside the heliosphere as a proxy for the ISM properties of the three M dwarf stars in their study. If we had used the ISM values around the Sun ($\nu_{\text{ISM}}=25.7\,\text{km~s}^{-1}$ and $n_{\text{ISM}}=0.1\,\text{cm}^{-3}$), GJ~436's astrosphere would be about 75\% larger. Therefore, the values of \citet{Herbst2020} could be different depending on the ISM properties around each star.

Another parameter that may vary is the LIS. In our work we used the LIS values from observations outside the heliosphere made by Voyager. Our assumption comes from the fact that the LIS conditions are believed to be similar throughout the Galactic disk \citep{Strong2007, Neronov2017, Prokhorov2018, Aharonian2020, Baghmanyan2020} and that it provides a homogeneous Galactic cosmic ray background. However, the spectrum of cosmic rays could have variations locally \citep{Baghmanyan2020}. Stars close to cosmic ray acceleration regions, such as supernova remnants, can show local variations in cosmic ray spectra and have a larger flux of cosmic rays \citep{fatuzzo2006}. According to \citet{Baghmanyan2020}, the LIS spectrum is representative of the cosmic ray spectrum throughout the the Galaxy, except within 100\,pc of a cosmic ray accelerator. Since GJ~436 is 10.14\,pc away from the solar system, and the Sun is not near any cosmic ray accelerator region, any variation of the LIS is not relevant in the context of this work.

Finally, the list of M dwarf stars which host a planet is quite large \citep{Vidotto2019} and it opens up the possibility to further investigate the propagation of Galactic cosmic rays through their astrospheres. Since we are most interested in possible habitable planets, a possible future work would be the study of Galactic cosmic rays propagation around M dwarfs with exoplanets in the habitable zone. In addition, M dwarfs with magnetic field measurements would also be good targets since it would give further constrains on the stellar wind properties. Knowing the ISM properties around the star is another important factor which was one of the strengths of choosing GJ~436 as a target.

Since M dwarfs can have strong magnetic fields and have a much closer-in habitable zone it is important to investigate the flux of stellar cosmic rays, which are energetic particles generated by the star. Also, the strong magnetic fields and higher levels of magnetic activity in M dwarfs indicate that they should be efficient at accelerating stellar cosmic rays. Potentially habitable planets would therefore be located closer to the stellar cosmic ray source. Stellar cosmic rays could play a more important role for close-in planets than Galactic cosmic rays \citep{Segura2010, Grenfell2012, Tabataba2016, Fraschetti2019, Scheucher2020, Rodgers2021} and should be further investigated in future work. In the context of GJ~436, because it is not very magnetically active, stellar cosmic rays may not be as important as for other stars with stronger activity.

\section{Conclusions}
\label{sec:conc}
In this paper we investigated the propagation of Galactic cosmic rays through the astrosphere of the planet-hosting M dwarf system GJ~436. Galactic cosmic ray fluxes are suppressed in an energy-dependent way as they travel through magnetised stellar winds, known as the modulation of Galactic cosmic rays. Our main goal was to calculate the intensity of Galactic cosmic rays in the habitable zone of GJ~436 and at the orbital distance of GJ~436\,b and compare it with observations at the Earth. For that, we used a 1D cosmic ray diffusive transport equation to model the modulation of Galactic cosmic rays, including spatial and momentum advection of Galactic cosmic rays by the stellar wind. Given that the stellar wind of GJ~436 is not well constrained, we used two Alfv\'{e}n-wave-driven stellar wind models from \citet{Mesquita2020} consisting of a magnetically-dominated wind (case A) and a thermally-dominated wind (case B). With this, we were also able to investigate how the stellar wind regime could affect the propagation of Galactic cosmic rays. 

The wind models selected for this work have similar velocity and magnetic field profiles but with mass-loss rates two orders of magnitude different. Because of this difference, one model (case B) produced an astrosphere ten times larger than the other model (case A). The two wind cases show different fluxes of Galactic cosmic rays for the same distance. This difference in fluxes, however, is not that large, being less than half an order of magnitude different between cases A and B for the same orbital distance. The difference in both cases only occurs for energies lower than 1\,GeV -- the fluxes are more similar for higher energies. 

For distances larger than 10\,au diffusion is the main physical process dominating the cosmic ray propagation, which leads to little modulation of cosmic rays at $\sim$\,GeV energies. For these distances, a larger astrosphere results in only slightly lower Galactic cosmic ray fluxes in comparison with a smaller one (comparing the same orbital distance). Between $\sim 0.2\,$au and 1\,au the fluxes for the two wind cases begin to converge.

At 0.2\,au the fluxes of Galactic cosmic rays are nearly identical for both wind setups. At this orbital distance, we noticed a change in behaviour in the propagation of cosmic rays. For distances larger than 0.2\,au, the velocity and magnetic field of the stellar winds are very similar, leading to similar levels of modulation (with slightly lower fluxes for the larger astrosphere system). On the other side, for distances smaller than 0.2\,au, the magnetically-dominated wind modulates the Galactic cosmic rays more due to a higher local wind velocity (which leads to a smaller advective timescale).

In the habitable zone of GJ~436 (0.2--0.4\,au) the flux of Galactic cosmic rays are comparable to the intensities observed at Earth and are approximately the same for both of our wind setups. We provide an analytical fit to our spectra in the habitable zone of the star in \Cref{eq:ffa}. This fit can be used to reproduce the Galactic cosmic ray spectrum found in our work. We also analysed the flux of Galactic around GJ~436\,b (0.028\,au) and found that both wind regimes show a strong modulation of cosmic rays, with values four orders of magnitude smaller than intensities observed at the present-day Earth. The thermally-dominated wind (case B) shows intensities of Galactic cosmic rays twice as high as magnetically-dominated wind (case A). The results found here could be further used to investigate the propagation of Galactic cosmic rays through the magnetosphere and atmosphere of GJ~436\,b.

The stellar wind properties such as magnetic field and velocity are very important for cosmic ray propagation. In the case of GJ~436, in particular, our lack of knowledge on the wind regime (more thermally-dominated versus more magnetically-dominated), and consequently on the astrosphere size, do not strongly affect the propagation of Galactic cosmic rays on this system (for our choice of diffusion coefficients).

\section*{Acknowledgements}
This project has received funding from the European Research Council (ERC) under the European Union's Horizon 2020 research and innovation programme (grant agreement No 817540, ASTROFLOW). The authors wish to acknowledge the SFI/HEA Irish Centre for High-End Computing (ICHEC) for the provision of computational facilities and support. The authors acknowledge funding from the Provost's PhD Project award. We thank the anonymous reviewer for their careful reading of our paper and constructive comments.

\section*{Data Availability}
The data described in this article will be shared on reasonable request to the corresponding author.




\bibliographystyle{mnras}
\bibliography{bib/reference.bib}





\bsp	
\label{lastpage}
\end{document}